\documentclass[12pt]{article}
\usepackage{amsmath,amssymb,epsfig}
\makeatletter \@addtoreset{equation}{section} \makeatother
\renewcommand{\theequation}{\thesection.\arabic{equation}}
\newcommand{\ba}{\begin{array}}
\newcommand{\ea}{\end{array}}
\newcommand{\beq}{\begin{equation}}
\newcommand{\eeq}{\end{equation}}
\newcommand{\bea}{\begin{eqnarray}}
\newcommand{\eea}{\end{eqnarray}}

\def\bce{\begin{center}}
\def\ece{\end{center}}

\def\nonu{\nonumber}

\def\be{\beta}

\def\eps6{{\displaystyle \mathop{\epsilon}^{6}}{}}

\def\nab6{{\displaystyle \mathop{\nabla}^{6}}{}}

\def\0{{\sst{(0)}}}
\def\1{{\sst{(1)}}}
\def\2{{\sst{(2)}}}
\def\3{{\sst{(3)}}}
\def\4{{\sst{(4)}}}
\def\5{{\sst{(5)}}}
\def\6{{\sst{(6)}}}
\def\7{{\sst{(7)}}}
\def\8{{\sst{(8)}}}

\def\ba{\begin{array}}
\def\ea{\end{array}}
\def\beq{\begin{equation}}
\def\eeq{\end{equation}}
\def\be{\begin{equation}}
\def\ee{\end{equation}}

\def\eps{\epsilon}

\def\ba{\begin{array}}
\def\ea{\end{array}}
\def\beq{\begin{equation}}
\def\eeq{\end{equation}}
\def\be{\begin{equation}}
\def\ee{\end{equation}}

\def\eps{\epsilon}

\newcommand{\bean}{\begin{eqnarray*}}
\newcommand{\eean}{\end{eqnarray*}}

\begin{document}
\thispagestyle{empty} \addtocounter{page}{-1}
\begin{flushright}
KIAS-P07008 \\
{\tt hep-th/0703015}\\
\end{flushright}

\vspace*{1.3cm}

\centerline{ \Large \bf Meta-Stable Brane Configuration and Gauged
  Flavor Symmetry }
\vspace*{1.5cm}
\centerline{{\bf Changhyun Ahn} 
} 
\vspace*{1.0cm} 
\centerline{\it 
Department of Physics, Kyungpook National University, Taegu
702-701, Korea} 
\vspace*{0.8cm} 
\centerline{\tt ahn@knu.ac.kr} 
\vskip2cm

\centerline{\bf Abstract}
\vspace*{0.5cm}

Starting from an ${\cal N}=1$ supersymmetric electric gauge 
theory with the gauge group $Sp(N_c) \times SO(2N_c')$
with fundamentals for the first gauge group factor and a
bifundamental, 
we apply
Seiberg dual to the symplectic gauge group only and arrive at the ${\cal
  N}=1$ supersymmetric  dual magnetic
 gauge theory 
with dual matters including the gauge singlets and superpotential.
By analyzing the F-term equations of the dual magnetic superpotential, we 
describe the intersecting brane configuration of type IIA string
theory corresponding to the meta-stable nonsupersymmetric vacua 
of this gauge theory.  

\baselineskip=18pt
\newpage
\renewcommand{\theequation}
{\arabic{section}\mbox{.}\arabic{equation}}

\section{Introduction}

The ${\cal N}=1$ $SU(N_c)$ gauge theory with $N_f$
fundamental flavors
has a vanishing superpotential before we deform this theory by mass 
term for
quarks.
When we add an adjoint field to this theory, then there exists a nonzero
superpotential for this adjoint field. On the other hand, when we add
a symmetric tensor flavor(or antisymmetric tensor flavor) to ${\cal N}=1$
$SU(N_c)$ SQCD with $N_f$ fundamental flavors, in general, there exists a
nonzero quartic superpotential consisting of these matter
fields.
In the type IIA brane configuration, this gauge theory can be described by three
NS5-branes, D6-branes,
D4-branes and an
orientifold 6-plane. The coefficient functions appearing the
above quartic superpotential depend on both how the two outer NS5-branes
are rotated with respect to a middle NS5-brane and how the D6-branes
are rotated with respect to a middle NS5-brane. Then by tuning these
two rotation angles in appropriate way, one can make the above nonzero
quartic superpotential to vanish.  
For the brane configuration 
description corresponding to ${\cal N}=1$ supersymmetric gauge theory, 
see the review paper \cite{GK}. 

Both deforming the electric gauge theory by adding the mass for the
quarks and taking the Seiberg dual magnetic theory from the electric 
theory are
necessary to find out meta-stable supersymmetry breaking vacua in the
context of dynamical supersymmetry breaking \cite{ISS,IS}.
When we take the Seiberg dual from the electric theory, the vanishing
superpotential in the electric theory 
makes it easier to deal with its nonvanishing dual
magnetic superpotential since the reduced superpotential in the
magnetic theory has simple form to analyze.   

On the other hand, 
the ${\cal N}=1$ product gauge group $SU(N_c) \times SU(N_c')$ with $N_f$
fundamentals, $N_f'$ fundamentals and two bifundamentals can be
described by similar expression for the superpotential 
to the one in the single gauge group in
previous paragraph. In general,
one can also think of a multiple NS5-branes for 
the outer NS5-branes together with a
single middle NS5-brane.
Since there is no orientifold 6-plane in this
case, compared with previous brane configuration, 
the rotations of outer NS5-branes and those of D6-branes with
respect to a middle NS5-brane are completely arbitrary. 
Therefore, the coefficient
functions in the superpotential 
depend on four rotation angles rather than two.
Then by tuning these
four rotation angles appropriately, one can make the nonzero
superpotential to vanish in the electric theory.   

Let us add an orientifold 4-plane in this brane configuration
for product gauge group with matter contents we mentioned 
in previous paragraph.
One expects that the gauge group will be a product gauge group between
a symplectic group with $O4^{+}$-plane and an orthogonal group with 
$O4^{-}$-plane. Under this orientifolding, the matter contents should
be changed appropriately: one kind of
fundamental flavors for each gauge group factor and one bifundamental.  
This particular brane configuration with $O4^{\pm}$-planes will play
the important role in this paper:
three
NS5-branes, D6-branes,
D4-branes and an
orientifold 4-plane.

A non-chiral model of dynamical supersymmetry breaking with stable 
nonsupersymmetric vacua was found in \cite{IT,IY} \footnote{It is
  called by ITIY model or IYIT model. }. By gauging the
``maximal'' subgroup of the
flavor symmetry group of this model, the existence of runaway behavior    
on the scalar potential at the region of the number of flavors
corresponding to \cite{IT,IY} was explained by using the brane
configuration \cite{dBHOO} in the magnetic theory where there are 
no D6-branes. 
Moreover, in \cite{Hashiba}, the different gauging of subgroup 
of the flavor symmetry group  \footnote{In \cite{Forste}, the diagonal
flavor group of ${\cal N}=1$ $SU(N_c)$ SQCD with massive flavors is
gauged and the additional color singlets are inserted both in an
electric theory and  magnetic theory. It is an open problem how to
realize this gauge theory from the type IIA brane configuration. On the
other hand, in \cite{CSST}, the flavor group is gauged but the Seiberg
dual is taken on this gauged flavor group rather than color group. } 
leads to two facts: for the nonabelian
gauged flavor group, the M5-brane configuration with massless quarks 
is nonholomorphic while
for the $U(1)$ gauged flavor group, supersymmetry breaking with stable 
nonsupersymmetric vacua with massless quarks involving
$t$-configuration 
exists by analyzing the corresponding
holomorphic M5-brane configuration.

In this paper, 
starting from an ${\cal N}=1$ supersymmetric electric gauge 
theory, 
we obtain the ${\cal
  N}=1$ supersymmetric  dual magnetic
 gauge theory(with dual matters and corresponding superpotential) 
which is the model of \cite{IT,IY} with ``gauged'' flavor symmetry group.
Based on this magnetic brane configuration, we 
describe the intersecting brane configuration of type IIA string
theory corresponding to the meta-stable nonsupersymmetric vacua 
of this gauge theory by recombination of color D4-branes and flavor
D4-branes and moving those D4-branes to (45) directions. 
 
In section 2, we review the type IIA brane configuration corresponding
to the electric theory based on the ${\cal N}=1$ $Sp(N_c) \times
SO(2N_c')$ 
gauge theory
with matter contents and deform this theory by adding the mass term
for the quarks. 
Then we construct the Seiberg dual magnetic theory which is 
${\cal N}=1$ $Sp(\widetilde{N}_c) \times SO(2N_c')$ gauge 
theory with corresponding dual
matters as well as various gauge singlets, by brane motion and linking
number counting. 

In section 3, we consider the nonsupersymmetric meta-stable
minimum by looking at the magnetic brane configuration we obtained in
section 2
and present 
the corresponding intersecting brane configuration of type IIA string
theory,  along the line of 
\cite{OO,FGU,BGHSS,Ahn07-1,Ahn07,Ahn06-1,Ahn06} and 
we describe M-theory lift of this supersymmetry breaking 
type IIA brane configuration.

In section 4, we make some comments on the future directions.

\section{The ${\cal N}=1$ supersymmetric electric and magnetic 
brane configuration}

In this section, we consider ${\cal N}=1$ $Sp(N_c)$ gauge theory with 
gauged flavor symmetry group $SO(2N_c')$ in electric theory first and
then we move on its dual magnetic gauge theory 
 ${\cal N}=1$ $Sp(\widetilde{N}_c)$ gauge theory with 
the same gauged flavor symmetry group $SO(2N_c')$. We present the corresponding
brane configurations in 
Figure 1 and Figure 2 respectively.
 
\subsection{Electric theory}

The gauge group of an electric theory \cite{Hashiba} is given by $Sp(N_c) 
\times SO(2N_c')$
and the matter contents are \footnote{By introducing the chiral multiplets 
$X_1$ and $X_2$ that are symmetric tensor in $Sp(N_c)$ and antisymmetric tensor
in $SO(2N_c')$
respectively, the $Sp(N_c) \times SO(2N_c')$ gauge theory in
different context can be
described by
the tree-level superpotential that is
given by
$
W = W_1(X_1) + W_2(X_2) +  Q_{12}^T X_1 Q_{12} +  Q_{12} X_2 Q_{12}^T
$ where $Q_{12}$ is a bi-fundamental and $W_i(X_i)$ is a polynomial
for the field $X_i$. See also the same gauge theory with different
matters in \cite{ILS} where there exist more meson fields in the
magnetic theory and see \cite{OO1} for unitary gauge group in the
quiver gauge theory. }

$\bullet$
$2N_f$ quark fields $Q$ is in the 
representation $({\bf 2N_c},{\bf 1})$ under  the gauge group

$\bullet$
The flavor-singlet field $X$ is in the
bifundamental representation $({\bf 2N_c},{\bf 2N_c'})$
under  the gauge group

The anomaly free global symmetry is given by $SU(2N_f) \times U(1)_R$ and
the strong coupling scales for $Sp(N_c)$  and $SO(2N_c')$ are denoted by 
$\Lambda_1$ and $\Lambda_2$ respectively.
The theory is asymptotically free when 
$3(2N_c+2)-2N_f-2N_c'> 0$ for the $Sp(N_c)$ gauge theory
and when $3(2N_c'-2)-2N_c > 0$ for the
$SO(2N_c')$
gauge theory by computing the coefficients of beta function. 

The type IIA brane configuration for this theory \cite{Hashiba} 
can be described by
$2N_c$ color 
D4-branes (01236) suspended between a middle NS5-brane (012345) 
and the right NS5'-brane (012389) along $x^6$
direction
together with $2N_f$ D6-branes(including mirrors) 
at values of $x^6$ that are between 
those corresponding to the positions of the middle NS5-brane and the
right NS5'-brane.
Moreover, an extra the left NS5'-brane is located at the left hand side of
a middle NS5-brane along the $x^6$ direction and there exist $2N_c'$
color D4-branes
suspended 
between them. Of course, there is an orientifold 4-plane (01236) which
acts as $(x^4,x^5,x^7,x^8,x^9) \rightarrow (-x^4,-x^5,-x^7,-x^8,-x^9)$.
The RR charge of O4-plane flips sign each time it passes through
NS5-brane or NS5'-brane \cite{EJS}. These are shown in Figure 1.

The classical superpotential by deforming this theory by adding the
mass term for the quarks is given by
\footnote{The mass matrix $m$ in (\ref{superpotential}) is 
  antisymmetric in the flavor indices and is given by $
-i \sigma^2 \otimes \mbox{diag}(m_1, m_2, \cdots, m_{N_f})$ where
$\sigma^2$ is a second Pauli matrix explicitly.
The quadratic term  in (\ref{superpotential}) is given by
$Q J Q$ explicitly where the color indices are contracted. Here
$J$ is a symplectic metric and antisymmetric in the color indices
and is given by $J=i \sigma^2
\otimes 
{\bf 1}_{N_c \times N_c}$ explicitly.
\label{J}} 
\bea
W = m Q Q
\label{superpotential}
\eea
and this brane configuration can be summarized as follows \footnote{We
introduce two complex coordinates $v \equiv x^4 + i x^5$ and $w \equiv
x^8 + i x^9$, as usual. }:

$\bullet$
One middle NS5-brane(012345)

$\bullet$ 
Two NS5'-branes(012389)

$\bullet$
$2N_f$ D6-branes(0123789) 

$\bullet$
$2N_c$ color D4-branes(01236)

$\bullet$ 
$2N_c'$ color D4-branes(01236)

$\bullet$ 
One O4-plane(01236)

Now we draw this electric brane configuration in Figure 1 which was
already found in  \cite{Hashiba} for the massless quarks and we put
the coincident D6-branes in the nonzero $v$ direction(and its mirrors).
The electric quarks $Q$ correspond to 4-4 strings connecting the
$N_c$ color D4-branes with $N_f$ flavor D4-branes(and its mirrors) 
where one can see these flavor D4-branes more clearly if move $N_f$
D6-branes to the left all the way past NS5-brane(and its mirrors)
and after that there exist newly created $N_f$ D4-branes connecting 
$N_f$ D6-branes
and NS5-brane in this process.  
The bifundamentals $X$ correspond to   4-4 strings connecting the
$N_c$ color D4-branes with $N_c'$ color D4-branes(and its mirrors) 
\footnote{This brane configuration can be obtained from the 
brane configuration
  corresponding to the product
gauge group $SU(N_c) \times SU(N_c')$ theory \cite{BH,BHKL} with different
matter contents. The superpotential, in
general, has the form of 
$W = m_1 X_1^2 + m_2 X_2^2 + X_1 \widetilde{F} F + X_2 \widetilde{F} F
+ Q X_1 \widetilde{Q} + Q' X_2 \widetilde{Q}'$.
The masses are
 given by $m_1 =\tan \theta_1$ and $m_2=\tan \theta_2$ where
 $\theta_i$ is an rotation angle \cite{Barbon} of outer 
$NS5_{\theta_i}$-brane with
 respect to a middle NS5-brane. This superpotential can be reduced 
after O4-plane orientifolding procedure
\cite{Tatar,Ahn97}. By integrating the adjoint fields $X_1$ and $X_2$, 
the superpotential is given by 
$W \sim \left( \frac{1}{m_1} + \frac{1}{m_2} \right) X^4$. Now it is
evident that for $\theta_1=\theta_2 =\frac{\pi}{2}$,
the superpotential vanishes as we observed above.}. 

\begin{figure}[ht]
   \epsfxsize=3in 
\centerline{\epsffile{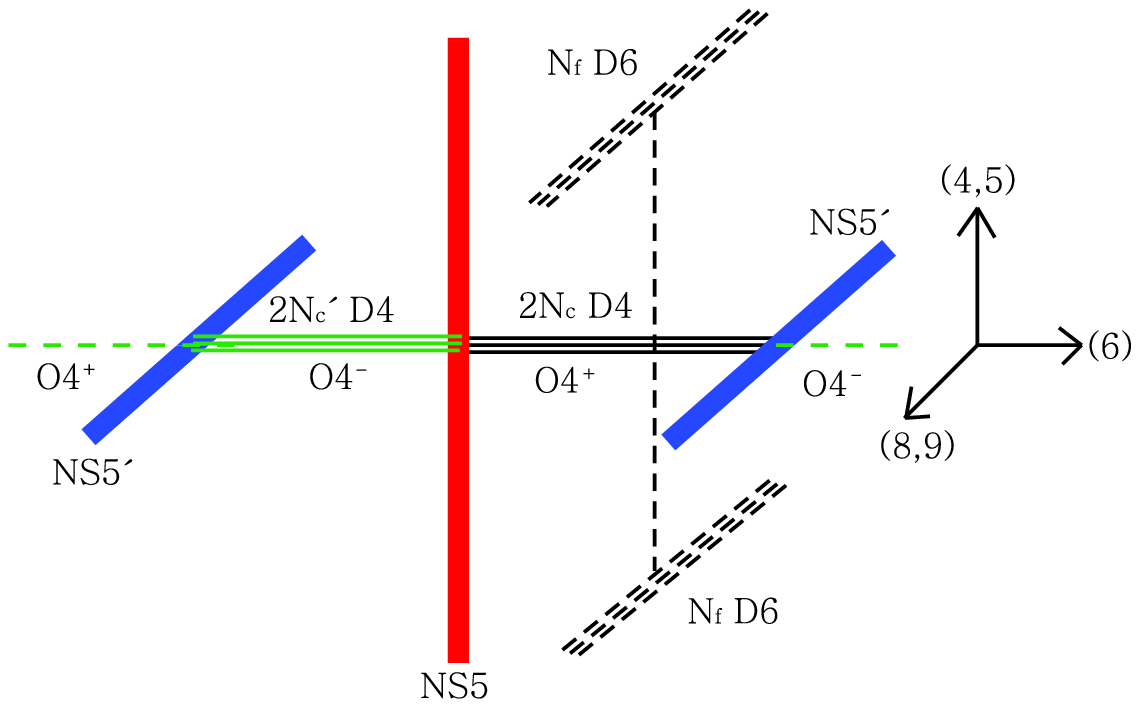}}
   \caption[FIG. \arabic{figure}.]{ 
The ${\cal N}=1$ 
supersymmetric electric brane configuration for the 
$Sp(N_c) \times SO(2N_c')$ gauge
theory  with 
$2N_f$ quark fields $Q$ in the fundamental
representation and 
 singlet field $X$  in the
bifundamental representation.
The origin of $(v,w)$ is located at the intersection between NS5-brane
and D4-branes. 
The D6-branes have nonzero $v$ coordinates where $v=\pm m$ for equal
massive case.
 }
\label{fig1}
\end{figure}

\subsection{Magnetic theory}

Let us take the Seiberg dual for the first gauge group factor $Sp(N_c)$ 
while keeping the second gauge group factor $SO(2N_c')$ untouched. 
Suppose that
$\Lambda_1 >> \Lambda_2$. This can be done by field theory side or 
type IIA string theory side via brane motion. 
Let us  do this by latter \cite{EGK,EGKRS,GK,Ahn07,Ahn07-1}.
After we move  a middle NS5-brane to the right all the way past
the right $NS5_R'$-brane, the linking number \cite{HW}
of NS5-brane from Figure 2 is given by 
$
L_5 =\frac{(2N_f)}{2}-1-(1)-2\widetilde{N}_c
$.
Note that $O4^{+}$-plane with RR charge $+1$ realizes a symplectic
gauge group while $O4^{-}$-plane with RR charge $-1$ does an
orthogonal gauge group. So we put $n_{4R}=-1$ due to the
$O4^{-}$-plane and  $n_{4L}=1$ due to the
$O4^{+}$-plane above as well as the contributions from the D4-branes.
Originally, it was 
$
L_5=-\frac{(2N_f)}{2}+ 1 -(-1)
+2N_c-2N_c'
$
from Figure 1 before the brane motion.
Also in this case, 
we put $n_{4R}=1$ due to the
$O4^{+}$-plane and  $n_{4L}=-1$ due to the
$O4^{-}$-plane.
Therefore, by the linking number conservation and equating these two
$L_5$'s each other, 
we are left with the number of colors in the magnetic
theory \cite{Hashiba}
\bea
\widetilde{N}_c = N_f + N_c'-N_c -2.
\nonu
\eea

Let us draw this magnetic brane configuration in Figure 2 which was
already found in  \cite{Hashiba} for the massless quarks and we put
the coincident D6-branes in the nonzero $v$ directions(and its mirrors).
During this process from an electric theory to the magnetic theory, 
the $N_f$ created D4-branes(and its mirrors) connecting between
D6-branes and NS5'-brane can move freely in the $w$ direction.
Moreover since $2N_c'$ D4-branes are suspending between two equal
$NS5'_{L,R}$-branes located at different $x^6$ coordinate, these D4-branes
can slide along the $w$ direction also.

\begin{figure}[ht]
   \epsfxsize=3in 
\centerline{\epsffile{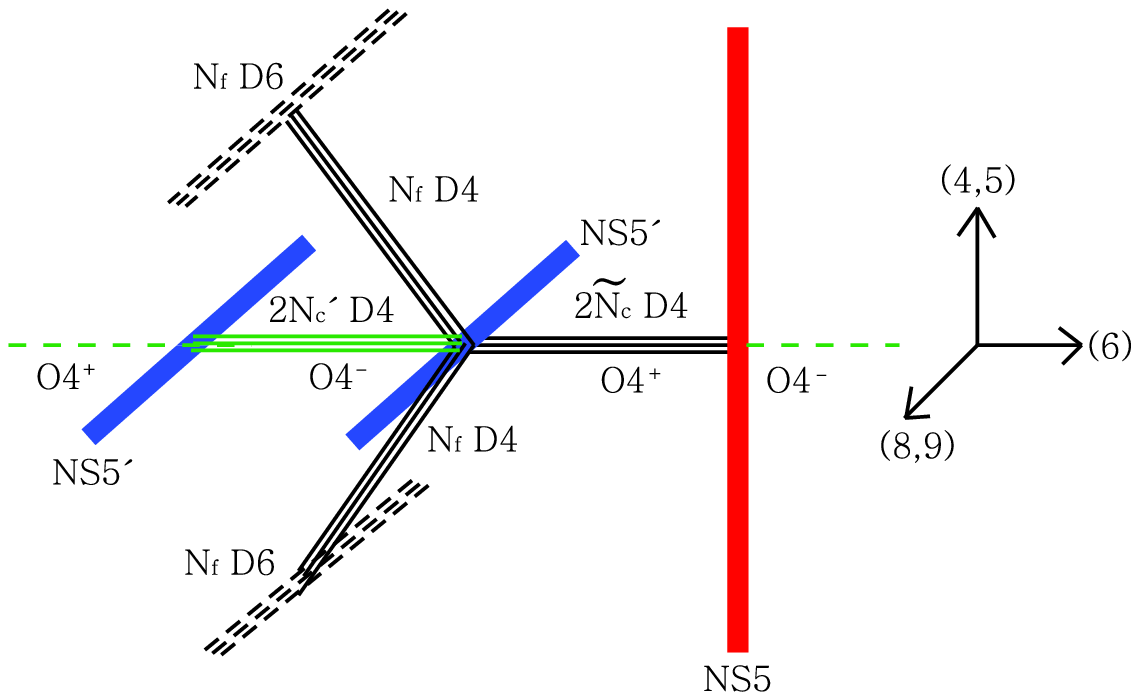}}
   \caption[FIG. \arabic{figure}.]{ 
The ${\cal N}=1$ 
supersymmetric magnetic brane configuration for the 
$Sp(\widetilde{N}_c= N_f + N_c'-N_c -2) \times SO(2N_c')$ gauge
theory  with 
$2N_f$ quark fields $\widetilde{Q}$,  
a singlet field $\widetilde{X}$,
 the gauge singlet $\Phi$, 
$2N_f$ fields $N$, and
 an antisymmetric two index tensor $S$.   }
\label{fig2}
\end{figure}

Then the weakly coupled dual magnetic gauge group \footnote{The ${\cal
  N}=1$ $Sp(\widetilde{N}_c)$ gauge theory with matter contents
  $\widetilde{Q}$ and $S$ and the superpotential $W_{dual}=
S \widetilde{Q} \widetilde{Q} + m S$ can be interpreted as a dual magnetic
   description of $Sp(N_c=N_f-\widetilde{N}_c-2)$ gauge theory
  with $2N_f$ massive quarks when the number of flavors satisfies $N_f
  > \widetilde{N}_c+2$. Then the Intriligator-Thomas-Izawa-Yanagida
  model(which can be also denoted as IYIT model in some literatures) 
\cite{Terning,Terning1,IT,IY}
can be obtained as an electric $Sp(N_c=-1)$ gauge theory with the condition
  $N_f=\widetilde{N}_c+1$ \cite{Hashiba}. For this particular number
  of flavor, there exists a quantum constraint $\mbox{Pf} \; S =
  \widetilde{\Lambda}^{2N_f}$ \cite{IP} and this leads to 
  the supersymmetric vacua when $m^{N_f}=\widetilde{\Lambda}^{2N_f}$ 
\cite{dBHOO,Hori} which corresponds to a supersymmetric M5-brane
  configuration. \label{conflict}} 
at the scale $E$ where $\Lambda_2
<< E$ is given
by $Sp(\widetilde{N}_c) \times SO(2N_c')$
and the matter contents are 

$\bullet$
$2N_f$ quark fields $\widetilde{Q}$ 
 in the 
$({\bf 2\widetilde{N}_c},{\bf 1})$ 

$\bullet$
the flavor-singlet field $\widetilde{X}$  in the
bifundamental representation $({\bf 2\widetilde{N}_c},{\bf 2N_c'})$

$\bullet$ the gauge-singlet $\Phi$ in the representation $({\bf
  1},{\bf N_c'(2N_c'-1)})$ 

$\bullet$
$2N_f$ fields $N$  in
$({\bf 1}, {\bf 2N_c'})$

$\bullet$ an antisymmetric two index tensor(in the flavor
indices) $S$ in the representation $({\bf
  1},{\bf 1})$ 

Magnetic quarks $\widetilde{Q}$ correspond to 4-4 strings connecting the
$\widetilde{N}_c$ color D4-branes with $N_f$ flavor D4-branes(and its
mirrors), $\widetilde{X}$ 
corresponds to  to 4-4 strings connecting the
$\widetilde{N}_c$ color D4-branes with $N_c'$ color 
D4-branes(and its mirrors), 
$\Phi$ is a flavor-singlet corresponding to
4-4 strings connecting two different color D4-branes among $N_c'$ 
color D4-branes(and its mirrors), $N$ 
corresponds to 4-4 strings connecting the
$N_c'$ color D4-branes with $N_f$ flavor D4-branes(and its mirrors)
and
$S$ corresponds to 
4-4 strings connecting two different flavor D4-branes among $N_f$
flavor 
D4-branes(and its mirrors).

In the dual theory, since 
there exist  $2N_f$ fundamental fields $\widetilde{Q}$ and  
one bifundamental field  
$\widetilde{X}$ which will give rise to the contribution of $2N_c'$,
the coefficient of the beta function \footnote{
The index of representation $R$ is given by 
$K(N_1)=2$, $K(\mbox{symm.})=2(N_1+2)$  
and $K(\mbox{adj}=\mbox{antisymm.})=2(N_1-2)$ for $SO(N_1)$ group and
$K(2N_2)=2$, $K(\mbox{antisymm.})=2(2N_2-2)$  and 
$K(\mbox{adj.}=\mbox{symm.})=2(2N_2+2)$ for $Sp(N_2)$ group.
Recall that the coefficient of the beta function is 
$b=\frac{1}{2} \left[3K(\mbox{adj.})-
\sum_{\mbox{matter}}K(R)\right]$.  }
is
$
b_{Sp(\widetilde{N}_c)}= 3(2\widetilde{N}_c+2)-2N_f-2N_c'
$
and since there are $2\widetilde{N}_c$ fundamental 
fields $\widetilde{X}$ for the 
flavor index of the first factor,
an antisymmetric tensor $\widetilde{\Phi}$ 
which will contribute to $(2N_c'-2)$ and $2N_f$ fields $N$, 
the coefficient of the beta function
 is
$
b_{SO(2N_c')} = 3(2N_c' -2)-2\widetilde{N}_c-(2N_c' -2)-2N_f$.
Therefore, both $Sp(\widetilde{N}_c)$ and 
$SO(2N_c')$ gauge couplings are IR free
by requiring the negativeness of the coefficients of beta function.
One can rely on the perturbative calculations at low energy 
for this magnetic IR free region $b_{Sp(\widetilde{N}_c)} < 0$ and 
$b_{SO(2N_c')} < 0$.

The dual magnetic  tree level superpotential, by adding the mass term
for the $Q$ in electric theory corresponding to add
a linear term in $S$ in dual magnetic theory \cite{Hashiba}, 
is given by
\bea
W_{dual} = \left( S \widetilde{Q} \widetilde{Q} + \widetilde{X} \Phi
\widetilde{X} + \widetilde{Q} N \widetilde{X} \right)
+ m S
\label{finalsuper}
\eea
where one can add intermediate scale  in the meson terms 
relating the strong coupling scale $\Lambda_1$ of the 
electric gauge group $Sp(N_c)$ to the scale of $\widetilde{\Lambda}_1$
of its magnetic dual gauge group $Sp(\widetilde{N}_c)$. 
Here the mesons are given, in terms of the fields defined in the
electric theory, 
by \footnote{The meson field $S$ is given by
$Q J Q$ where the color indices are contracted. Here
$J$ is given by the footnote \ref{J}. 
Similarly, the meson field 
$\Phi$ is given by
$X J X$ and 
the meson field $N$ is $Q J X$ where the color indices are contracted. }
\bea
S \equiv Q Q, \qquad \Phi \equiv  X X, \qquad N \equiv  Q X.
\nonu
\eea
The fluctuations of the singlet $S$ correspond to the motion in the
$(789)$ directions of the flavor D4-branes in Figure 2. Similarly,
the fluctuations of the singlet $\Phi$ correspond to the motion in the
$(789)$ directions of the D4-branes suspended two NS5'-branes in
Figure 2.
Here $\widetilde{Q}$ is fundamental for
the gauge group index and antifundamental for the flavor index.
Then, $\widetilde{Q} \widetilde{Q}$ has rank $2\widetilde{N}_c$ while $m$ has a
rank $2N_f$.  Therefore, the F-term condition, the derivative the 
superpotential $W_{dual}$ with respect to $S$, cannot be satisfied 
if the rank $2N_f$ exceeds $2\widetilde{N}_c$. 
This is so-called rank condition and the supersymmetry is broken.    

The classical moduli space of vacua can be obtained from F-term
equations.
From the F-terms $F_{\widetilde{Q}}$ and $F_{S}$, one gets
$S \widetilde{Q} + N \widetilde{X} =0=   \widetilde{Q} \widetilde{Q} +
m$.
Similarly, one obtains 
$\Phi \widetilde{X} + \widetilde{Q} N=0 = \widetilde{X} \widetilde{X}$ from 
the F-terms $F_{\widetilde{X}}$ and $F_{\Phi}$. 
Moreover, there is a relation 
$ \widetilde{Q} \widetilde{X}=0$ from the F-term $F_{N}$.
Then, one obtains 
the following solutions 
\bea
< \widetilde{Q}> =  \left(
\begin{array}{c}
i \sqrt{m}  {\bf 1}_{2\widetilde{N}_c}  \\
0
\end{array}
\right),  \quad
<S>  =
 \left(
\begin{array}{cc}
0  & 0 
 \\
0 & \Phi_0  {\bf 1}_{(N_f-\widetilde{N}_c)} \otimes i \sigma^2 
\end{array}
\right), \quad
<\widetilde{X}>=0= <N>
\label{vacuum}
\eea
where $\Phi_0  {\bf 1}_{(N_f-\widetilde{N}_c)} \otimes i \sigma^2$ 
is an arbitrary 
$2(N_f-\widetilde{N}_c) \times 2(N_f-\widetilde{N}_c)$ antisymmetric 
matrix and the zeros of 
$<  \widetilde{Q}>$  are $2(N_f-\widetilde{N}_c) \times 2\widetilde{N}_c $
zero matrices. 
Similarly, the zeros
of $2N_f \times 2N_f$  matrix $S$ are assumed also.
Then $\Phi_0$ and $i \sqrt{m}$
parametrize
a pseudo-moduli space.
Let us expand around on a point on (\ref{vacuum}) as done in \cite{ISS}. 
That is, 
\bea
\widetilde{Q}   = 
\left(
\begin{array}{c}
i \sqrt{m}  {\bf 1}_{2\widetilde{N}_c} +
(\delta \chi_{A} + \delta \chi_{S})
 {\bf 1}_{\widetilde{N}_c} \otimes i \sigma^2 \nonu \\
\delta \varphi
\end{array}
\right), 
\qquad
S =
 \left(
\begin{array}{cc}
\delta Y  & \delta Z^T
 \\
-\delta Z & \Phi_0  {\bf 1}_{(N_f-\widetilde{N}_c)} \otimes i \sigma^2
\end{array}
\right)
\nonu
\eea
as well as the fluctuations for $N$ and $\widetilde{X}$.
Then the superpotential becomes 
\bea
W_{dual}^{fluct} & = &  \Phi_0 \left( \delta \varphi \;  
 \delta \varphi  + m \right) +
  \delta Z^T \; \delta \varphi   \; i \sqrt{m}
+  \delta Z \; i \sqrt{m}  \; 
\delta \varphi  
\nonu \\
 & + & \left( 
\delta Y \; \delta \chi_{A}  \; i \sqrt{m}  
+ \cdots \right)
+ \mbox{(cubic)}
\label{s}
\eea
where $\mbox{(cubic)}$ stands for the terms that are cubic or higher
in the fluctuations and $\cdots$ contains some parts from the second
and third
terms in (\ref{finalsuper}) that are not relvant:
There are two kinds of terms, 
the vacuum of $<\widetilde{Q}>$ multiplied by 
$\delta N \delta \widetilde{X}$ and 
the vacuum of $<\Phi>$ multiplied by $\delta \widetilde{X} 
\delta \widetilde{X}$.
By redefining these as 
$\delta \hat{N} \delta \hat{\widetilde{X}}$ and $\delta
\hat{\widetilde{X}}
\delta \hat{\widetilde{X}}$ respectively, they do not enter the 
contributions for the one loop result. 
Then to quadratic order, the model splits into two sectors where the
first piece given by the first line of (\ref{s}) is an O'Raifeartaigh type model 
and the  second piece given by the second line of (\ref{s})
is supersymmetric and will not contribute to the
supertrace. The fields 
$\delta \chi_{A,S}$ and $\delta Y$
couple to the supersymmetry breaking fields $\delta \varphi$  
via terms of cubic and higher
 order in the fluctuations.
Then, the remaining relevant terms of superpotential are given by
the first line of 
(\ref{s}).
At one loop, the effective potential $V_{eff}^{(1)}$ for $\Phi_0$ 
can be obtained 
from this superpotential which consists of 
the matrices $M$ and $N$ of \cite{Shih} 
where the defining function ${\cal F}(v^2)$ can be
computed.
Using the equation (2.14) of \cite{Shih} 
of $m_{\Phi_0}^2$ and ${\cal F}(v^2)$, one gets 
that $m_{\Phi_0}^2$ will contain $(\log 4 -1) > 0$.
This implies that these vacua are stable.

Note that in the original model of \cite{IT,IY}, when $N_f
=\widetilde{N}_c +1$, the result of \cite{CLP} implies that when
the extremization of a tree-level superpotential with massless quarks   
conflicts with a quantum constraint, as observed in the footnote
\ref{conflict}, 
the low energy 
effective theory near the origin of moduli space is an O'Raifeartaigh
model \cite{O'Raifeartaigh} and  
the sign of mass squared for the pseudoflat direction at
the origin is calculable and there is a local minimum around the
origin. In this case the $U(1)_R$ symmetry is preserved. On the other hand, 
the effective theory
becomes noncalcuable for large moduli space.    
For the discussion of $Sp(1)\sim SU(2)$ SQCD with two flavors, see 
\cite{Terning,Terning1,MN}.   

Remember that in the original model of \cite{IT,IY}, the global flavor
group is given by $SU(2\widetilde{N}_c+2)$ and in \cite{Hashiba} the
$SO(2N_c')$ subgroup of $SU(2\widetilde{N}_c+2)$ for massless quarks 
was gauged where $1
\leq N_c' \leq \widetilde{N}_c$ while the ``maximal'' subgroup 
$SO(2\widetilde{N}_c+2)$ of $SU(2\widetilde{N}_c+2)$ was gauged in 
\cite{dBHOO}. 
The above magnetic brane configuration can be also obtained 
directly from ${\cal N}=1$ $Sp(\widetilde{N}_c)$ SQCD with $N_f$ 
fundamentals(let us focus on the massless case) 
by replacing $2N_c'$ D6-branes with an $NS5'_L$-brane(attaching the same
number $2N_c'$ D4-branes along the $x^6$ direction) and then the
remaining $2(N_f-N_c')$ D6-branes can be shifted to the new $2N_f$
D6-branes
which is consistent with the global symmetry group 
$SU(2N_f)$ \cite{Hashiba}. The appropriate consideration for the
O4-plane 
should be included.   

\section{Nonsupersymmetric meta-stable brane configuration
\label{four} }

Now we recombine $\widetilde{N}_c$ D4-branes among $N_f$ D4-branes 
connecting between D6-branes and $NS5_R'$-brane with those 
connecting between $NS5'_R$-brane and NS5-brane(and its mirrors) and push
them in $v$ direction from Figure 2 \cite{Ahn06,Ahn06-1,Ahn07,Ahn07-1}. 
Of course their mirrors will move
to $-v$ direction in a ${\bf Z}_2$ symmetric manner due to the 
$O4^{+}$-plane. 
After this procedure, there are no color D4-branes between 
$NS5'_R$-brane and NS5-brane.
For the flavor D4-branes, we are left with only 
$(N_f-\widetilde{N}_c)$ D4-branes(and its mirrors).  

Then the minimal energy supersymmetry breaking brane configuration is
shown in Figure 3.
If we ignore the left $NS5'_L$-brane, $2N_c'$ D4-branes and $O4^{+}$-plane
connecting to the left $NS5'_L$-brane(detaching these from Figure 3), 
then this brane configuration 
corresponds to  the minimal energy supersymmetry breaking brane
configuration
for the ${\cal N}=1$ SQCD with the magnetic gauge group 
$Sp(\widetilde{N}_c)$ with
$N_f$ massive flavors \cite{FGU,Ahn06-1}.

\begin{figure}[ht]
   \epsfxsize=3in 
\centerline{\epsffile{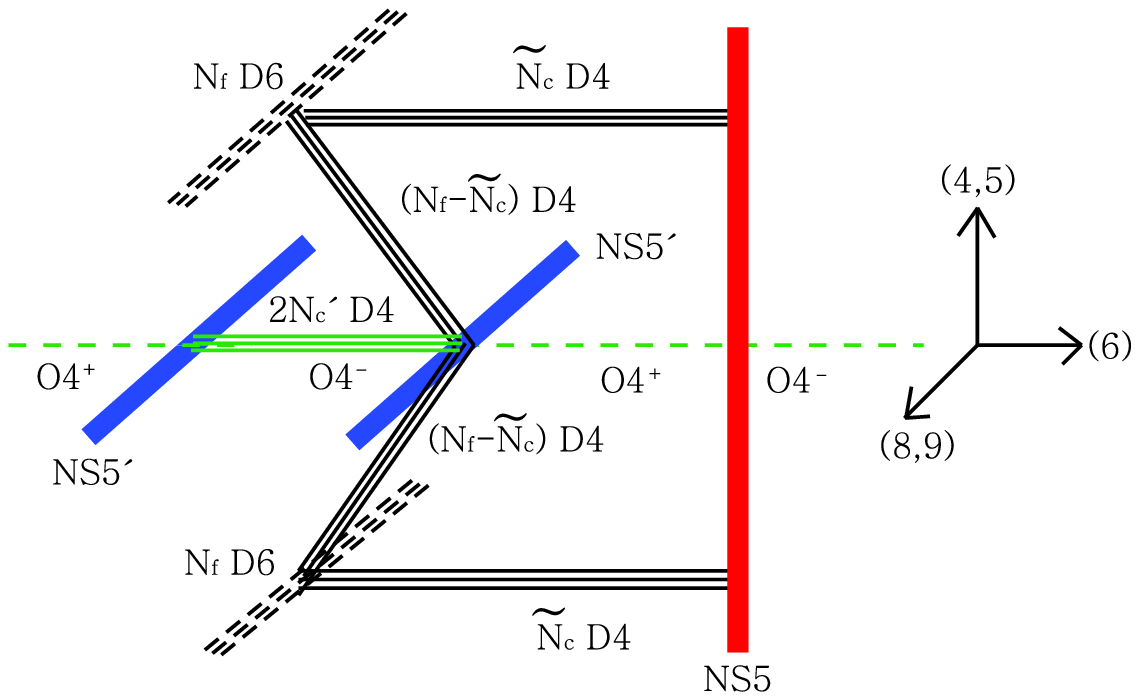}}
   \caption[FIG. \arabic{figure}.]{ 
The nonsupersymmetric minimal energy
brane configuration for 
 the 
$Sp(\widetilde{N}_c= N_f + N_c'-N_c -2) \times SO(2N_c')$ gauge
theory  with 
$2N_f$ quark fields $\widetilde{Q}$,  
a singlet field $\widetilde{X}$,
 the gauge singlet $\Phi$, 
$2N_f$ fields $N$, and
 an antisymmetric two index tensor $S$.
For the M-theory lift,  
the position of these $NS5'_{L,R}$-branes are given by $v=m$ and those for the
D4-branes is given by $v=2m$(its mirrors are given by $v=0$).  }
\label{fig3}
\end{figure}

The type IIA and M-theory brane construction for the ${\cal N}=2$
gauge theory was described by \cite{LLL97} and after lifting 
the type IIA description we explained so far to M-theory, the 
corresponding magnetic M5-brane configuration with equal mass for the
quarks where the gauge group is given by 
$Sp(\widetilde{N}_c) \times SO(2N_c')$ with fundamental matters for
the first gauge group factor, in a background space of $x t = 
\prod_{k=1}^{N_f} (v^2 -e_k^2)$ where this four dimensional space
replaces (45610) directions,
is characterized by 
\bea
t^3 - ( v^{2\widetilde{N}_c +2} + m^{N_f} + \cdots ) t^2 + 
( v^{2N_c'} + \cdots) t - v^2  
\left[\prod_{k=1}^{N_f} (v^2 -e_k^2 ) \right] =0
\label{curve}
\eea
where $e_k$ is the position of the D6-branes in the $v$ direction and
we ignored the lower power terms in $v$ in $t^2$ and $t$ 
denoted by $\cdots$
and the scales for the gauge groups in front of the first term and the
last term, for simplicity
\footnote{Of course, the corresponding supersymmetric 
electric M5-brane configuration from
Figure 1 can be written similarly
and the cubic equation can be written as follows:
$ 
t^3 - (v^{2N_c+2} + m^{N_f} + \cdots ) t^2 + 
( v^{2N_c'} + \cdots) \left[\prod_{k=1}^{N_f} (v^2 -e_k^2 )\right] t - v^2  
\left[\prod_{k=1}^{N_f} (v^2 -e_k^2 ) \right]^2 =0$
where $e_k$ is the position of the D6-branes in the $v$ direction and
we put the equal mass $m$ for the quarks. The relevant part of
\cite{LLL97} is explained in the subsection 4.4 where we need to
specify when their function
$J_s \equiv \prod_{k=i_{s-1}+1}^{i_s} (v^2 -e_{k}^2)$ where 
$i_{k}$ is related to the number of D6-branes between 
the $(k-1)$-th and $k$-th NS5-brane by  
$d_{k}= 
i_{k}-i_{k-1}$
is nontrivial. In the present electric case, the D6-branes are
located between NS5-brane and $NS5_R'$-brane from Figure 1. It is easy
to see that $J_1$ is a function of $v^2$ and their $J_0$ and $J_2$ are
constant.}.
When we consider the gauging of the
maximal subgroup of the
flavor symmetry group \cite{dBHOO}, the corresponding M5-brane configuration 
can be read off from this by putting the contribution from
D6-branes(the coefficient of $v^2$ in the last term) to 1.
Also the massless case in \cite{Hashiba} can be obtained from this 
by taking the massless limit. Moreover, the case without D6-branes can
be obtained from this curve by appropriate limit and will become
(3.33) of \cite{LLL97}.
If we take the limit where the last term in (\ref{curve}) vanishes,
then the remaining piece becomes ${\cal N}=1$ $Sp(\widetilde{N}_c)$
SQCD with $2N_c'$ matter fields \cite{LLL97,AOT} and similarly for the
limit where the first term of (\ref{curve}) 
will vanish, we obtain the remaining ${\cal
N}=1$ $SO(2N_c')$ SQCD with matter fields \cite{LLL97,AOT1}.  

From this curve (\ref{curve}) 
of cubic equation for $t$ above, the asymptotic regions 
can be classified by looking at 
the first two terms providing NS5-brane asymptotic region, 
next two terms providing $NS5_R'$-brane asymptotic region and 
the final two terms giving $NS5_L'$-brane asymptotic region
as follows:
 
1. $v \rightarrow \infty$ limit implies
\bea
w \rightarrow 0, \quad y \sim    v^{2\widetilde{N}_c+2} + \cdots \quad
\mbox{NS asymptotic region}.   
\nonu
\eea

2.  $w \rightarrow \infty$ limit implies
\bea
v & \rightarrow &   m, \quad 
y \sim  w^{2N_f-2N_c'+2}
 +\cdots
\quad \mbox{$NS_{L}'$ asymptotic region}, 
\nonu
\\
v & \rightarrow &  m, \quad  
y \sim w^{2N_c'-2\widetilde{N}_c-2}
+\cdots
\quad \mbox{$NS_{R}'$ asymptotic region}. 
\nonu
\eea

As observed in \cite{Ahn06-1}, the two $NS5'_{L,R}$-branes are moving in the
$v$ direction holding everything else fixed instead of moving
D6-branes in the $v$ direction. Then the mirrors of D4-branes are
moved appropriately. In Figure 1, the origin of $v$ coordinate is
located at the intersection of NS5-brane and D4-branes while 
in the coordinate we describe above $NS_{L,R}'$ asymptotic region, the
origin is located at the position of mirror D4-branes. Then the
location of the original D6-branes is given by $v=2m$. In other words,
there is a shift in $v$ direction.
The harmonic function sourced by the D6-branes can be written
explicitly by summing of two contributions from two(original and its
mirrors) 
$N_f$ D6-branes, as in \cite{Ahn06-1} 
\bea
V(s) = 1  + \frac{N_f R}{
  \sqrt{f(s)^2 + s^2}} + \frac{N_f R}{\sqrt{(f(s)- 2\Delta x)^2 + s^2}}.
\nonu
\eea
Note that when we compare with the result of \cite{BGHSS},
the last term above is an extra contribution. 
For the straight line solution where $x^4=f(s) = \Delta x$, the
differential equation for $g(s)$, which is equal to the magnitude of 
$w$, can be solved and it leads to
\bea
g(s) \sim \exp \left[ \frac{s}{4(N_c+1)R} \right] \left( \frac{s+
    \sqrt{(\Delta x)^2+ s^2}}{R} \right)^{\frac{N_f}{2(N_c+1)}}
\nonu
\eea
and the $s$-independent integration constant can be fixed by the
boundary condition from the above classification 2. 
From this solution, it is easy to see that 
even if $\Delta x$ goes to zero, 
the function $g(s)$ does not vanish. There is no smooth nonholomorphic
M-theory curve.
So the extra piece in the potential does not remove 
an instability from a new M5-brane mode.

\section{Conclusions and outlook}

In this paper, by  gauging the maximal subgroup 
of the global flavor group for the model \cite{IT,IY}, 
we have 
described the intersecting brane configuration of type IIA string
theory in Figure 3 corresponding to the meta-stable nonsupersymmetric vacua 
of this gauge theory.

Also it is straightforward to take 
the dual gauge theory $Sp(N_c) \times SO(2\widetilde{N}_c')$
and construct a meta-stable brane configuration in IIA string theory.

There exist many other examples of dynamical supersymmetry breaking 
in \cite{ADS}-\cite{LT} and in \cite{Terning,Terning1,IS,SS}. 
So it is natural to ask whether these examples(or their
generalizations
in many different directions) 
have the type IIA string theory brane configuration(if not, one should find
out this brane configuration first) and it is an open
problem to find out any new meta-stable brane configuration, along the
lines of this paper if any.
Some of the gauge theories can be described by 
type IIB string theory where there exist D5-branes, 
NS5-branes, NS5'-branes, D7-branes and O7-planes.  
We need to perform the
Seiberg dual and add nonzero mass term for the quarks and sometimes the vanishing
superpotential in the electric theory 
makes it easier frequently to deal with its nonvanishing dual
magnetic superpotential. 

\vspace{.7cm}

\centerline{\bf Acknowledgments}

I would like to thank Y. Ookouchi for discussions on ISS model \cite{ISS} 
and its relevant subjects.
This work was supported by grant No.
R01-2006-000-10965-0 from the Basic Research Program of the Korea
Science \& Engineering Foundation.  
I thank KIAS(Korea Institute for 
Advanced Study) for hospitality  where
this work was undertaken.

\end{document}